%% file: main.tex
\documentclass[a4paper,twoside]{article}

\usepackage{epsfig}
\usepackage{subcaption}
\usepackage{calc}
\usepackage{amsthm}
\usepackage{multicol}
\usepackage{pslatex}
\usepackage{apalike}
\usepackage[bottom]{footmisc}
\usepackage{amssymb}
\usepackage{amsthm}
\usepackage{amsmath}
\usepackage{stmaryrd}
\usepackage{listings}
\usepackage{multicol}
\usepackage{tikz}
\usepackage{mathtools}
\usepackage{pgfplots}
\usepackage{csvsimple}
\usepackage[linesnumbered,algoruled]{algorithm2e}
\SetKwProg{Fn}{Function}{ is}{end}

\usepackage{color, colortbl}
\usepackage{bbold}

\usetikzlibrary{matrix, positioning}
\usetikzlibrary{shapes,arrows}
\usepackage{graphicx} %package pour l'insertion d'images
\usepackage{todonotes}
\usepackage{SCITEPRESS}     % Please add other packages that you may need BEFORE the SCITEPRESS.sty package.

\newtheorem{defi}{Definition}[section]
\newtheorem{theo}{Theoreme}[section]

\newtheorem{lem}{Lemme}[section]

\newtheorem{example}{Example}[subsection]

\begin{document}

\title{Near-collisions and their Impact on Biometric Security}
\author{\authorname{DURBET Axel\sup{1}, GROLLEMUND Paul-Marie\sup{2}, LAFOURCADE Pascal\sup{1} and THIRY-ATIGHEHCHI Kevin\sup{1}}
  \affiliation{\sup{1}Université Clermont-Auvergne, CNRS, Mines de Saint-Étienne, LIMOS, France}
  \affiliation{\sup{2}Université Clermont-Auvergne, CNRS, LMBP, France}
  %\email{\{f\_author, s\_author\}@ips.xyz.edu, t\_author@dc.mu.edu}
}


\abstract{ Biometric recognition encompasses two operating modes.
  The first one is biometric identification which consists in determining the identity of an
  individual based on her biometrics and requires browsing the entire database (\textit{i.e.}, a $1$:$N$ search).
  The other one is biometric authentication which corresponds to verifying claimed biometrics of an individual (\textit{i.e.}, a $1$:$1$ search)
  to authenticate her, or grant her access to some services. The matching
  process is based on the similarities between a fresh and an enrolled biometric
  template.
  Considering the case of binary templates, we investigate how a highly populated database
  yields near-collisions, impacting the security of both the operating modes. Insight into the security of binary
  templates is given by establishing a lower bound on the size of templates and an upper bound on the size of a template
  database depending on security parameters. We provide efficient algorithms for partitioning a leaked template database
  in order to improve the generation of a master-template-set that can impersonates any enrolled user and possibly some future users.
  Practical impacts of proposed algorithms are finally emphasized with experimental studies.
}

\onecolumn \maketitle \normalsize \setcounter{footnote}{0} \vfill

\input{Introduction}

\input{Background}

\input{Meth_decoup}

\input{Create_mp}
\input{Attack_eval}

\input{Conclusion}

\section*{Acknowledgement}
The authors acknowledges the support of the French Agence Nationale de la Recherche (ANR), under grant ANR-20-CE39-0005 (project PRIVABIO).

\bibliographystyle{apalike}
{\small
  \bibliography{biblio}}

\input{Annexes}

\end{document}

%% file: Introduction.tex
\section{Introduction}

With the continuous growth of biometric sensor markets, the use of biometrics is
becoming increasingly widespread. Biometric technologies provide an
effective and user-friendly means of authentication or identification through
the rapid measurements of physical or
behavioral human characteristics.
For biometric identification and authentication schemes, biometric
templates of users are registered with the system.
The first operating mode consists in determining the identity of an
individual based on similarity scores calculated from all the
enrolled templates and the fresh provided template. The latter corresponds
to the verification of the claimed identity based on a similarity score
calculated from the assigned
enrolled template and a fresh template.
As a consequence,
service providers need to
manage biometric databases in a manner similar to managing
password databases.

The leak of biometric databases is more dramatic since, unlike passwords, biometric data serve as long term identifiers and cannot be easily revoked. The consequences of stolen biometric templates are impersonation attacks and the compromise of privacy. Essential
security and performance criteria that must be met by biometric recognition systems are identified in ISO/IEC 2474~\cite{ISO24745} and ISO/IEC 30136~\cite{ISO30136}: \emph{Irreversibility}, \emph{unlinkability}, \emph{revocability} and \emph{performance preservation}.

Biometric templates are generated from biometric measurements (\textit{e.g.}, a
fingerprint image). They result from a chain of treatments,
an extraction of the features (\textit{e.g.}, using Gabor
filtering~\cite{ManMa96,JaPrHoPa2000}) followed eventually by a Scale-then-Round process~\cite{AKK20} to
accommodate better handled representations, \textit{i.e.,} binary or
integer-valued vectors. These templates are then protected either through
their mere encryption, or using a Biometric Template Protection (BTP), \emph{e.g.,} a cancelable biometric transformation such as Biohashing~\cite{JIN20042245,LUMINI2007} or any other salting method. For more details on BTP
schemes, the reader is referred to the surveys \cite{Survey-2015,Survey-2016,patel2015cancelable}. The use
of a BTP scheme is in general preferred since its goal is to address the aforementioned criteria. However, note that cancelable biometric transformations are prone to inversion attacks, at least in the sense of second-preimages~\cite{durbet2021authentication}. They even lead sometimes to the compromise of privacy with a good approximation of a feature vector~\cite{LaChRo13,GKLT20}.

Recent works have also demonstrated that recognition systems are vulnerable to dictionary attacks based on master-feature vectors~\cite{RoMeRo17,BoRoToMeRo18}. A master-feature vector
is a set of synthetic feature vectors that can %fortuitously
match with a large number of other feature vectors. This can naturally be extended to the problem of generating master-templates and masterkeys. The notion of masterkey has recently been addressed in~\cite{gernot2021biometric} to produce backdoors with the aim of implementing biometric-based access rights. In the same topic, the present paper analyses the security of biometric databases by making some recommandations, and by proposing attacks using the notions of master-template, master-feature and masterkey.

%and can be stored either 
%from biometric cryptosystems (fuzzy vault, fuzzy sketches, fuzzy extractors) 
%or from

\paragraph*{\textbf{Contributions.}}

Our main contribution is an efficient partitioning
algorithm which accelerates attacks aiming to generate
master-key or master-feature vector. % and define a security bound over the size of database.
Numerical studies on implementations of the proposed algorithm show a reduction of the computational time by a factor of up to
$38$ in certain settings.
In addition, we show a link with the closest string problem with an arbitrary number of words, for which we provide a solution
using Simulated ANNealing (SANN).
% Formal definitions of master-template, $\epsilon$-covering template
% and, some technical terms and concepts are introduced.
% Moreover, a limit on the number of templates in a database is determined by using near-collisions.
Moreover, we determine a bound on the size of a database in function of the template space dimension and the decision threshold,
thus preventing near-collisions with a high probability. Specifically, for a secure database, the recommanded template size is $n=512$ bits with a threshold of the order of $10$\% of $n$, \emph{i.e.,} around $50$ bits. Setting these paramaters in this way rules out attacks based on master-templates and ensures a good recognition accuracy.
Finally, some indications are provided for handling basic database operations
such as
addition or deletion of users.

%{\color{blue}Rmqs :
%
%        - Il y a un souci d'ordre dans la description de l'organisation.
%
%        - Dans les papiers de conf (au moins sécu), on fait référence à des
%        sections et sous-sections, mais pas des parties. Certains écrivent ``section''
%        quelle que soit le niveau de la section...
%
%        - Attention aux accords avec la 3ème personne du singulier. Je pense les avoir
%        corrigés.}

%{\color{red}
\paragraph*{\textbf{Outline.}}
% Formal definitions of master-template~\ref{EMT}, $\epsilon$-covering-template~\ref{ECT}
% and, some technical terms and concepts are introduced
% in Section~\ref{BackGround}.
% Algorithm~\ref{SpacePartitioningAlgorithm} which provides a partition
% of a database is developed in Section~\ref{SPM}.
% The experimentation of Algorithm~\ref{SpacePartitioningAlgorithm} are in
% Section~\ref{AttackEval} and compared to
% a naive Algorithm~\ref{Gluttony} in the Appendices.
% To provide Algorithm~\ref{SpacePartitioningAlgorithm}, solving the closest string problem~\ref{CSP} was necessary. Thus,
% we develop a solution in Section~\ref{FMT}.
% In Section~\ref{GC}, we show how to use Algorithm~\ref{SpacePartitioningAlgorithm} to improve the computation of masterkey-set~\ref{MKS} and
% master-feature-set~\ref{MFS} defined in Section~\ref{BackGround}.
% Section~\ref{SecureParamK} describes how near-collision\ref{NCol}
% can be used to define a secure parameter $k$ which depends on the
% template space dimension and a threshold. In addition,
% some example of such a $k$ are provided in Table~\ref{ChooseK}. Some indications are provided for handling
% basic database operations such as addition or deletion of users in Section~\ref{CS:ADD} and ~\ref{CS:REM}.
In Section~\ref{BackGround}, we introduce some notations, background material as well as definitions of new notions such as master-template and $\epsilon$-covering-template. In Section~\ref{SPM}, we describe an algorithm which provides a segmentation of a database in order to focus on potential master-templates.
%This section is divided into several subsections which all deal with a different part of the Algorithm~\ref{SpacePartitioningAlgorithm}. First, in Section~\ref{AC}, we talk about the clustering aspect of our algorithm.
% In particular in Section~\ref{FMT}, we develop how to solve the near string problem, which is also a part of the proposed algorithm provided in Section~\ref{DPA}.
In Section~\ref{FMPCBS}, we show  how this algorithm can be used to improve the computation of masterkey-set and of the
master-feature-set. Moreover, we describe how near-collisions
can be used to define a secure parameter $k$ which depends on the
template space dimension and a threshold.
% Section~\ref{SecureParamK} describes how near-collision\ref{NCol}
% can be used to define a secure parameter $k$ which depends on the
% template space dimension and a threshold.
% This secure parameter is a countermeasure to the attacks provided in Section~\ref{GC}.
We also explain why the secure parameter is a countermeasure
% Afterwards, in Section~\ref{CS:ADD} and \ref{CS:REM},
and, the case of a user which is added or removed from the database are studied.
In Section~\ref{AttackEval}, we provide some experimentations
in order to assess the performance of the proposed algorithm and to detail in practice how the near string problem is solved.
%In Section~\ref{SPEval}, we give the result of tests over the algorithm~\ref{SpacePartitioningAlgorithm} and
%in Section~\ref{TestalgoPaul} our method to solve the near string problem is tested.
%}

% \bigskip
% 
% Commentaire : Je ne sais pas si une partie contribution serait nécessaire (personnellement je ne pense pas mais cela peut dépendre de la revue), mais si c'est le cas, il faut reformuler la partie entre [A] et [B], notamment en y ajoutant un peu plus d'explications.

%% file: Background.tex
\section{Preliminaries}\label{BackGround}

% Je ferai un retour sur cette partie une fois qu'elle sera finalisée. 
% Sur la demande de Kevan, je me concentre sur les parties qui me concernent plus.

A biometric system is a method of authentication or identification
based on biometric data.  The main idea is to transform the biometric
data into a template to match the four aforementioned criteria, i.e.,
irreversible, unlinkable, revocable and performance preservation. It must be able to compare
template and determine if they belong to the same person. The template
is constructed by combining a feature vector derived from the
biometric data and a secret parameter named token which can be for
example a password.  A biometric authentication or identification
system always starts by using a \emph{feature extraction scheme} to
extract some information from the biometric image to construct a
feature vector~\cite{Rathasheme}.  A database partitioning method can
be applied to each biometric system for this.
In this paper, we focus on templates expressed as binary vectors, but the results below can be adapted to every template representations.
%In this article, we focus on those which have their template represented by binary vector. 
%Nevertheless, it can be extended to every representation of templates.

% The present work requires some definitions~\ref{FES} and~\ref{CBTS} 
%from~\cite{GKLT20}.

In the following, we let $(\mathcal{M}_I,\operatorname{Dist}_I)$, $(\mathcal{M}_F,\operatorname{Dist}_F)$ and
$(\mathcal{M}_T,\operatorname{Dist}_T)$ be three metric spaces, where
$\mathcal{M}_I$, $\mathcal{M}_F$ and
$\mathcal{M}_T$ represent the image space,
the feature space and the template space, respectively;
and $\operatorname{Dist}_I$, $\operatorname{Dist}_F$ and $\operatorname{Dist}_T$ are
the respective distance functions. Note that $\operatorname{Dist}_I$ and $\operatorname{Dist}_F$ are instantiated
with the Euclidean distance, while $\operatorname{Dist}_T$ is instantiated with the Hamming
distance.

\begin{defi}[Feature extraction scheme]
    \label{FES}
    A biometric \emph{feature extraction scheme} is a pair of deterministic polynomial time algorithms $\Pi:=( E,V)$, where:
    \begin{itemize}
        \item $E$ is the feature extractor of the system, that takes biometric data
              $I \in \mathcal{M}_I$ as input, and returns a feature vector $F \in \mathcal{M}_F$.
        \item $V$ is the verifier of the system, that takes two feature vectors
              $F=E(I)$, ${F^\prime}=E(I')$, and a threshold $\tau_F$ as input, and
              returns $True$ if $\operatorname{Dist}_F(F, {F^\prime}) \leq \tau_F$, and returns $False$ if
              $\operatorname{Dist}_F(F, {F^\prime}) > \tau_F$.
    \end{itemize}
\end{defi}

For the sake of privacy, biometric data (the feature vector) should be designed in a such way that it prevents information leakage. This motivates the use of a cancelable
biometric transformation scheme.

\begin{defi}[Cancelable biometric transformation scheme]
    \label{CBTS}
    Let $\mathcal{K}$ be the token (seed) space, representing the
    set of tokens to be assigned to users. A \emph{cancelable biometric scheme} is
    a pair of deterministic polynomial time algorithms $\Xi:=(\mathcal{T}, \mathcal{V})$, where:
    \begin{itemize}
        % \item $\mathcal{G}$ is the secret parameter generator of the system, that takes a token (seed) $s\in \mathcal{K}$ as input, and returns a secret parameter set $sp$.
        \item $\mathcal{T}$ is the transformation of the system, that takes a
              feature vector $F \in \mathcal{M}_F$ and the token parameter $P\in \mathcal{K}$ as input,
              and returns a biometric template $T=\mathcal{T}(P,F) \in \mathcal{M}_T$.
        \item $\mathcal{V}$ is the verifier of the system, that takes two biometric
              templates $T$ = $\mathcal{T}(P,F)$,
              ${T^\prime}=\mathcal{T}({P^\prime},{F^\prime})$, and a
              threshold $\tau_T$ as input; and returns
              $True$ if $\operatorname{Dist}_T(T, {T^\prime}) \leq \tau_T$, and
              returns $False$ if $\operatorname{Dist}_T(T, {T^\prime}) > \tau_T$.
    \end{itemize}
\end{defi}

In this paper, the template space is, unless otherwise specified,
$\mathbb{F}_2^n = \left(\mathbb{Z}\big / 2\mathbb{Z}\right)^n$, equipped with
the Hamming distance denoted by $d_H$.  As the template space is a
metric space, we denote it as $\left(\mathbb{F}_2^n,d_H\right)$.  In
our case, the verifier is the Hamming distance, but the transformation
does not need to be specified. As we work on a set of template, we
denote it as  \emph{Template DataBase} (\emph{TDB}).

\begin{defi}[Template database or TDB]
    Let $\left(\Omega,d\right)$ be the template sp\-ace equipped with
    the distance $d$.  A subset $L \subset \Omega$ such that $L \not =
        \emptyset$ and $L \not = \Omega$ is a \emph{template database
        (TDB)}, or just a database.
\end{defi}

As with hash functions, an antecedent of a transform can be searched in order to steal
a password or a pass tests using this hash function.
This preimage can be the exact feature vector or a nearby preimage.

\begin{defi}[Template preimage]
    Let $I\in \mathcal{M}_I$ be a biometric image,
    and $T=\Xi.\mathcal{T}(P,\Pi.E(I))\in\mathcal{M}_T$ for some secret parameter
    $P$. A \emph{template preimage} of $T$ with respect to $P$
    is a biometric image $I^*$ such that
    $T=\Xi.\mathcal{T}(P,\Pi.E(I^*))$.
\end{defi}

\begin{defi}[Nearby template preimage]
    Let $I\in \mathcal{M}_I$ be a biometric image, a threshold $\epsilon_B$,
    and $T=\Xi.\mathcal{T}(P,\Pi.E(I))\in\mathcal{M}_T$ for some secret parameter
    $P$. A \emph{nearby template  preimage} of $T$ with respect to $P$
    is a biometric image $I^*$ such that
    $d(T,\Xi.\mathcal{T}(P,\Pi.E(I^*))) < \epsilon_B$.
\end{defi}

%{\color{red} Reformuler le paragraphe suivant.}

% Our goal is to find a masterkey-set. The main idea is to create a list
% of passwords as a masterkey. As the notions as been introduced
% vector. Hence, we introduce the adapted following
% in~\cite{gernot2021biometric} but describe a master-feature
% definitions.
%~\ref{MK},~\ref{MKS},~\ref{MF} and~\ref{MFS}.

The goal of an attacker can be to create a masterkey-set.
This is a set of tokens that allow to build all the templates
of a targeted database using %preferably 
the same feature vector.

\begin{defi}[Masterkey]
    \label{MK}
    Let $D = \left\lbrace v_i\right\rbrace_{i=1,\dots,n}$ be a template database where $v_i := \Xi.\mathcal{T}(x_i,s_i)$
    generated with distinct tokens $S = \left\lbrace s_i \right\rbrace_{i=1,\dots,n}$ and
    distinct biometric features $X = \left\lbrace x_i \right\rbrace_{i=1,\dots,n}$, and let $\tau_B$ be
    a threshold. Then, $m$ is a masterkey for $D$, with respect to $\tau_B$, if $\forall i \in
        \left\llbracket 1,n \right\rrbracket, \Xi.\mathcal{V}(\Xi.\mathcal{T}(x_i,m),\Xi.\mathcal{T}(x_i,s_i),\tau_B) = True$.
\end{defi}
Furthermore, in this context another objective of an attacker can be to find a masterkey-set or a master-feature-set, which are defined below.
\begin{defi}[Masterkey-set]
    \label{MKS}
    Let $D = \left\lbrace v_i\right\rbrace_{i=1,\dots,n}$ be a template database where $v_i := \Xi.\mathcal{T}(x_i,s_i)$
    generated with distinct tokens $S = \left\lbrace s_i \right\rbrace_{i=1,\dots,n}$ and
    distinct biometric features $X = \left\lbrace x_i \right\rbrace_{i=1,\dots,n}$, and let $\tau_B$
    a threshold. The set $D$ is said \emph{covered by a set of $r$ masterkeys} $\left\lbrace k_1,\dots,k_r\right\rbrace$ with respect to
    $\tau_B$, if $\forall i \in \left\llbracket 1,n \right\rrbracket, \exists j \in \left\llbracket 1,r \right\rrbracket$
    such that $\Xi.\mathcal{V}(\Xi.\mathcal{T}(x_i,k_j),\Xi.\mathcal{T}(x_i,s_i),\tau_B) = True$.
\end{defi}

Another goal of an attacker can be to create a master-feature-set.
This is a set of feature that allow to build all the templates
of a targeted database using preferably the same token.

\begin{defi}[Master-feature]
    \label{MF}
    Let $D = \left\lbrace v_i\right\rbrace_{i=1,\dots,n}$ be a template database where $v_i := \Xi.\mathcal{T}(x_i,s_i)$
    generated with distinct tokens $S = \left\lbrace s_i \right\rbrace_{i=1,\dots,n}$ and
    distinct biometric features $X = \left\lbrace x_i \right\rbrace_{i=1,\dots,n}$, and let $\tau_B$
    a threshold. Then, $m$ is a \emph{master-feature} for $D$, with respect to $\tau_B$, if $\forall i \in
        \left\llbracket 1,n \right\rrbracket, \Xi.\mathcal{V}(\Xi.\mathcal{T}(m,s_i),\Xi.\mathcal{T}(x_i,s_i),$ $ \tau_ B ) = True$.
\end{defi}

\begin{defi}[Master-feature-set]
    \label{MFS}
    Let $D = \left\lbrace v_i\right\rbrace_{i=1,\dots,n}$ be a template data\-base where $v_i := \Xi.\mathcal{T}(x_i,s_i)$
    generated with distinct tokens $S = \left\lbrace s_i \right\rbrace_{i=1,\dots,n}$ and
    distinct biometric features $X = \left\lbrace x_i \right\rbrace_{i=1,\dots,n}$, and let $\tau_B$
    a threshold. The set $D$ is said \emph{covered by a set of $r$ master-features} $\left\lbrace k_1,\dots,k_r\right\rbrace$ with respect to
    $\tau_B$, if $\forall i \in \left\llbracket 1,n \right\rrbracket, \exists j \in \left\llbracket 1,r \right\rrbracket$
    such that $\Xi.\mathcal{V}(\Xi.\mathcal{T}(k_j,s_i),\Xi.\mathcal{T}(x_i,s_i),\tau_B) = True$.
\end{defi}

%\begin{defi}[$\epsilon$-covering masterkey]
%A cancelable biometric database $D = \left\lbrace u_i\right\rbrace_{i=1,\dots,n}$ is said $\epsilon$-covered
%by a masterkey $x$, with $0 < \epsilon \leq 1$ if there exists a subset $D'$ of $D$ of $\epsilon n$ persons such
%that $x$ is a masterkey for $D'$. The optimal coverage percentage of $D$ is the maximal
%number $E$ such that $D$ is $E$-covered by a masterkey $x \in \mathcal{K}$.
%\end{defi}

Targeting random template to find a masterkey are often not efficient,
thus, to maximize the efficiency of the research of a masterkey-set,
we suggest to focus on $\epsilon$-covering templates.

\begin{defi}[$\epsilon$-cover-template]
    \label{ECT}
    Let $D$ be a template database and a distance $d$.
    An \emph{$\epsilon$-cover-template} of $D$ is $x$ such that
    $d(x,a) \leq \epsilon, \forall a \in D$.
\end{defi}
Note that, there are cases for which there is no possible $\epsilon$-cover-template.
\begin{example}
    \label{Exemp}
    Let $\epsilon = 1$, $n = 3$ and
    $D = \left\lbrace (0,0,0);(0,1,1);(1,0,1);(1,1,0) \right\rbrace \subset \mathbb{F}_2^3$.
    In this case, we have $\forall a,b \in D, d_H(a,b) \leq 2$, but there is no template $x$ such that
    $\forall a \in D, d_H(x,a) \leq 1$. The $1$-cover template does not exist.
    But, if we remove $(0,0,0)$ from $D$ then $x=(1,1,1)$ is an $1$-cover template
    for $D$.
\end{example}

As the $\epsilon$-covering template is non-unique, we also consider $\epsilon$-covering template-sets.
\begin{defi}[$\epsilon$-cover-template-set]
    \label{ECTS}
    Let $D$ be a template database and $d$ a distance.
    An \emph{$\epsilon$-cover-template-set} of $D$ is $\mathcal{C}$ such that
    $\forall u \in D$ and $\forall x~\in~\mathcal{C}$, $d(u,x) \leq \epsilon$.
\end{defi}
To construct a partition of a template database, we introduce strong and weak notions of $\epsilon$-master-template.
\begin{defi}[$\epsilon$-master-template or $\epsilon$-MT]
    \label{EMT}
    Let $\left(\Omega,d\right)$ be the template sp\-ace and $D$ a
    template database.  A template $t\in \Omega$ is an
    \emph{$\epsilon$-master-template }if $\forall t' \in D, d(t,t') \leq
        \epsilon$.
\end{defi}
\begin{defi}[$\epsilon$-master-template-set or $\epsilon$-MTS]
    Let $\left(\Omega,d\right)$ be the template space and $D$ a template database. A subset
    $T \subset \Omega$ is an \emph{$\epsilon$-master-template-set} if $\forall t' \in D, \exists t \in T$
    such that $d(t,t') \leq \epsilon$.
\end{defi}
Note that an $\epsilon$-master-template-set is a non-empty set: $D$ is
an $\epsilon$-master-template-set of itself but an
$\epsilon$-master-template of $D$ could be empty. Moreover, an
$\epsilon$-cover-template is an $\epsilon$-master-template and an
$\epsilon$-master-template-set is a set of $\epsilon$-cover-templates
which are not in the same $\epsilon$-cover-template-set.  We define a
near-collision and more precisely multiple-near-collision.

\begin{defi}[Near collision]
    \label{NCol}
    Let $\left(\Omega,d\right)$ be the template space and a
    threshold $\epsilon$ . There exists a \emph{near-collision} if $\exists a,b \in \Omega \mid d(a,b) \leq
        \epsilon$.
\end{defi}

\begin{defi}[$m$-near-collision]
    \label{MNCol}
    Let $\left(\Omega,d\right)$ be the template space and  a
    threshold $\epsilon$. There exists an \emph{$m$-near-collision} if $\exists a_1,\dots,a_m \in \Omega$
    such that
    $\forall i$ and $j \in \lbrace 1,\dots,m \rbrace, d(a_i,a_j) \leq \epsilon$.
\end{defi}
Thus, the search of an $\epsilon$-cover-template of $D$ a database
corresponds to the search of an at least $|D|$-near-collision for
which each template of $D$ is related to the collision.
% There exist
% two types of near-collision.  
% The near-collision is the ability
% to find two templates in the same ball.  
% While the strong is the
% ability to find a template in the same ball as another fixed in
% advance.

%% file: Meth_decoup.tex
\section{Database Partitioning}
\label{SPM}

The aim of this part is to determine the
smallest $\epsilon$-covering-template-set for a given database $D$.
%We propose a theoretical framework step by step, in order to deduce an algorithm.

% {\color{red}
%     (A mettre dans la conclusion/discussion.)
% 
%     The main idea is given a method which from a template, gives either the feature vector
%     or the password, the attacker wants to find this information for each person in a
%     minimum of steps. To minimize this number of steps, we propose to find a set of master-template
%     and attack it instead of all templates.
%     The formal framework of the utility of this method
%     is developed in the part~\ref{GC}.}
% 
\subsection{Agglomerative Clustering}
\label{AC}

Consider $M_D$ the dissimilarity matrix of
a template database $D$, for the Hamming distance.
The dissimilarity matrix $M_D$ is used to compute template clusters, denoted by
$C_{\epsilon}$, for which  the distance between two templates in the same cluster is at
most $s$. To perform this clustering, we use the agglomerative clustering
method which is a type of the hierarchical clustering.
This method consists in successively agglomerating the two closest groups of templates.
It begins with $|D|$ groups, one for each template, and it terminates when all the groups are merged as a unique one.

A standard post-processing is required to define at which iteration
the algorithm should be terminated so that a relevant
set of template clusters is obtained.
However, we define a termination condition so that
the clustering algorithm stop when it is not possible anymore to
obtain templates cluster verifying the following required property:
$\forall i \in \left\llbracket 1,n \right\rrbracket,$ $\forall a, b \in
    C_i, max(d_H(a,b)) \leq s$. The Agglomerative Clustering algorithm we
used then corresponds to a slight variation of the HACCLINK
(Hierarchical Agglomerative Clustering Complete LINK) presented
in~\cite{10.1093/comjnl/20.4.364}.

By using the aforementioned clustering method, we obtain a set of template clusters, for which the inner-cluster distance suggests
that it could exist at least one master-template for these templates.
An additional step is described below whose aim is to determine potential master-templates, if there exists some.

% We propose the algorithm~\ref{CCA} to perform the computation of clusters. This algorithm takes as
% input $D$ a template database and $s$ an integer and returns $Cls$ a partition of $D$ such that
% $\forall a,b \in C_i, max(d_H(a,b)) \leq s$. Clustering algorithms return labels and not
% clusters so, we need to organise clusters and construct the partition.
% 
% 
% \begin{algorithm}
%     \caption{Compute cluster algorithm}
%     \label{CCA}
%     \KwIn{$D,s$}
%     \KwOut{$Cls$}
%     Compute $\mathcal{M}_D$ the dissimilarity matrix of $D$. \\
%     Set $label$ to the result of the agglomerative clustering algorithm
%     with maximum distance allowed $s$. \\
%     Set $Cls$ to the partition constructed with $D$ and $label$.\\
%     \Return{$Cls$.}\\
% \end{algorithm}

\subsection{Master-Template of a Template Group}
\label{FMT}

We consider having a group of templates verifying  $\forall i \in \left\llbracket 1,n \right\rrbracket,$
$ \forall a,b \in C_i,$ $ max(d_H(a,b)) \leq s$, and for which we aim at finding a master-template.
We emphasize that this problem can be formulated as a modified case of closest-string problem which is defined as follows.
\begin{defi}[Modified closest-string problem]
    \label{MCSP}
    Given $S = \lbrace s_1,s_2,\dots,s_m \rbrace$ a set of strings with length $n$ and $d$ a distance, find
    a center string $t$ of length $m$ such that for every string $s$ in $S$, $d_H(s,t) \leq d$.
\end{defi}
The closest-string problem is known as an $NP$-hard
problem~\cite{Frances1997}, and there exist algorithms to solve that
kind of problem, see among others~\cite{CSPIP,CSPextr}.

\begin{defi}[Closest-string problem]
    \label{CSP}
    Given $S = \lbrace s_1,s_2,\dots,s_m \rbrace$ a set of strings with length $n$, find
    a center string $t$ of length $m$ minimizing $d$ such that for every string $s$ in $S$, $d_H(s,t) \leq d$.
\end{defi}

According to the link between both problems defined in Definition~\ref{CSP} and~\ref{MCSP}, we
can establish that the issue addressed in this paper is a hard problem,
which is specified in the following theorem,  whose proof is given in the Appendix~\ref{MYNPHARDProof}.

\begin{theo}[MCSP is NP-hard]
    \label{MYNPHARD}
    The modified closest-string problem is $NP$-hard.
\end{theo}
To the best of our knowledge, this problem has not been addressed in
the literature, then we propose an algorithm to solve it.  Moreover,
with regards to Theorem~\ref{MYNPHARD}, we deem that relying on brute
force type algorithm could not be efficient and that more parsimonious
algorithm must be investigated, notably stochastic algorithms.
However, more efficient upcoming methods could replace this part
without affecting the remainder of the database partitioning method
proposed in Section~\ref{SPM}.

We consider $D = \lbrace v_1,\dots,v_k\rbrace$ be a template database and $\mathcal{C}$ the
$\epsilon$-cover-template-set for $D$. The approach described below provides a constructive definition
of the elements of $\mathcal{C}$, if $\mathcal{C} \not = \emptyset$.
In particular, the following result emphasizes the link between $\mathcal C$
and the balls $B_i = \lbrace u \in \mathbb{F}_2^n | d_H(u,v_i) \leq \epsilon \rbrace$.
% Thus, $\mathcal{C} = \cap_{i \in \lbrace 0,\dots,k\rbrace} B_i $.
The proof is given in Appendix~\ref{Intertheoproof}.

\begin{theo}[$\mathcal{C}$ is the intersection of the balls of radius $\epsilon$]
    \label{Intertheo}
    Let $D = $ $\lbrace v_1,$ $\dots,v_k\rbrace$ be a template database and $\mathcal{C}$ the
    $\epsilon$-cover-template-set for $D$. \\ Then,  $\mathcal{C} = \cap_{i \in \lbrace 1,\dots,k\rbrace} B_i $.
\end{theo}

We denote by $p \in \mathcal C$ a master-template, and Theorem~\ref{Intertheo} indicates that determining all
the master-template $p$ reduces to determining the intersection of $k$
Hamming balls, which turns out to be formulated as the solutions of the following system:
%\begin{eqnarray}
%    \label{Sysys}
%    \begin{cases}
%        d_{H}(p,v_1) & \leq \epsilon \\
%        \vdots                       \\
%        d_{H}(p,v_k) & \leq \epsilon \\
%    \end{cases}
%\end{eqnarray}
\begin{equation}
    \label{Sysys}
    d_{H}(p,v_i) \leq \epsilon, \quad \forall i \in \{ 1,\dots,k\}.\\
\end{equation}
Notice that System~\ref{Sysys} is a linear system, hence we can rely on a binary ILP (Integer Linear Programming) to solve it and then to compute $\mathcal{C}$.
% Let two binary templates $x=(x_1, \ldots, x_n)$ and $y=(y_1,\ldots,y_n)$.
% The Hamming distance $d_{H}$ between $x$ and $y$ can be rewritten as an inner product between
% $(n+2)$-dimensional vectors in $\mathbb{Z}$, denoted $X$ and $Y$. We have
% \begin{eqnarray*}
%     d_{H}(x,y) & = & \sum_{i=1}^n x_i + \sum_{i=1}^n y_i - 2 \sum_{i=1}^n x_iy_i\\
%     & = & \langle X, Y \rangle
% \end{eqnarray*}
% where
% $$X=(x_1,x_2,\ldots,x_n,1,\sum_{i=1}^n x_i)$$
% and
% $$Y=(-2y_1,-2y_2,\ldots,-2y_n,\sum_{i=1}^n y_i,1).$$
%
%
%
% Equivalently, the system to solve is
% \begin{equation}
%     
%     M\dot X \leq E,
% \end{equation}
% 
% with the following notations:
% \begin{eqnarray*}
%     \begin{cases}
%         E^T = (\underbrace{\epsilon,\dots,\epsilon}_k)                                                                            \\
%         X^T = (\underbrace{p_1,\dots,p_n}_p,1,\sum_{i=1}^n p_i)                                                                   \\
%         Y_i = -2(\underbrace{(a_i)_1,\dots,(a_i)_n}_{a_i},-\dfrac{1}{2}\sum_{j=1}^n (a_i)_j,-\dfrac{1}{2}), & \forall i \in [1,k] \\
%         M = \left(\begin{array}{c}
%                 Y_1    \\
%                 \vdots \\
%                 Y_k    \\
%             \end{array}\right)
%     \end{cases}
% \end{eqnarray*}

However, solving this system could be time-consuming in real world cases since
there are as many parameters as the length of $p$, i.e., the dimension
$n$ of $\mathbb F_2^n$.
Therefore, we suggest reducing System~\ref{Sysys} by removing dependent variables.
To do so, the following necessary notations are introduced:
\begin{itemize}
    \item For $K = \lbrace k_1, \dots, k_{| K |}\rbrace \subset \lbrace 1,\dots,n \rbrace$, the Hamming distance
          over $K$ is denoted by:
          $\forall u,v \in \mathbb{F}_2^n,d_K = d_H((u_{k_1},\dots,u_{k_{| K |}}),(v_{k_1},\dots,v_{k_{| K |}}))$.
    \item Let $\mathcal{P}_D(K)$ a statement about $K \subset \lbrace 1,\dots,n \rbrace$. We said that $\mathcal{P}_D(K)$ holds
          if $\forall u,v \in D, d_K(u,v)
              \in \left\{0,|K|\right\}
              %= \begin{cases}
              %        | K | \\
              %        0
              %\end{cases}
          $.
          Examples of subsets $K$ for which $\mathcal{P}_D(K)$ holds or not are provided in Appendix~\ref{EXFMT}.
          % (tu pourras faire un 
          % exemple de chaque, avec à chaque fois un ensemble $D$. Il faudra réutiliser ces
          % deux exemples pour illustrer $I$.)
    \item Consider $I$ the smallest partition $\lbrace (K_1,\dots,K_{\mid I \mid}), K_i \subset  \left\lbrace 1, \dots,n \right\rbrace \mid \forall i \in \left\lbrace 1, \dots, \mid I \mid \right\rbrace \rbrace$ such that $\mathcal{P}_D(K_i)$  holds  for all $i \in \lbrace 1, \dots, n\rbrace$.
          As $I$ is the smallest possible partition, it allows us to reduce the dimension of System~\ref{Sysys} as much as it is possible.
          % An example of a partition $I$ is provided in Appendix~\ref{EXFMT}.
          %\item For $a \in D$, $A_K(a)$ denote $\lbrace b \in D \mid d_K(a,b) = 0 \rbrace \subset D$.
    \item For $p \in \mathbb{F}_2^n$ and $v \in D$, $n_{v,i}$ denotes $d_{K_i}(p,v)$ and $n^I_v$ denotes the parameters vector
          $(n_{v,1},\dots,n_{v,|I|})$, written $N = (n_1,\dots,n_{\mid I \mid})$ for short
          when the context is clear.
    \item The vector of distances $\left(d_H(v_1,v),\dots,d_H(v_{\mid D \mid},v)\right)$ is denoted by $d(v)$ with $v\in D$ and $D = (v_1,\dots,v_{\mid D \mid})$.
          %{\color{red}(mettre bien des $v_k$ à la place des $a_k$ partout, ou inversément.)}
\end{itemize}
Then, with these notations, Theorem~\ref{TheoPaul} can be established as follows, specifying a smaller version of System~\ref{Sysys}.
\begin{theo}
    \label{TheoPaul}
    For a given template database $D$ and for a given $v \in D$, consider $L = \lbrace p \in \mathbb{F}_2^n \, | \,
        AN \leq \epsilon - d(v)\rbrace$ with $N = n^I_v$,
    $\epsilon  = (\epsilon,\dots,\epsilon)^T$,$n_{v,i}$ denotes $d_{K_i}(p,v)$,
    $n^I_v$ denotes the parameters vector $(n_{v,1},\dots,n_{v,|I|})$
    and $A=(a_{i,j})$ a matrix of size $|I|\times |D|$
    whose the $(i,j)^\text{th}$ element is $$a_{i,j} = \begin{cases}
            1  & \text{if } d_{K_j}(v_1,v_i) = 0     \\
            -1 & \text{if } d_{K_j}(v_1,v_i) = |K_j| \\
        \end{cases}$$
    Then, $L = \mathcal{C}$ the
    $\epsilon$-cover-template-set for $D$.
\end{theo}
Proof is detailed in Appendix~\ref{TheoPaulProof}.

As $I$ is required to reduce System~\ref{Sysys}, we assure with Lemma~\ref{LemmaPaul} that $I \not = \emptyset$, whatever the configuration of the set $D$ is.
\begin{lem}[$I$ is not empty]
    \label{LemmaPaul}
    $\forall D \subset \mathbb{F}_2^n$ such that $\mid D \mid > 1$, $I \not = \emptyset$.
\end{lem}
Proof is given in Appendix~\ref{LemmaPaulProof}.
In the same vein, one can determine that $|I|\leq n$.
As $|I|$ corresponds to the number of parameters, the system described in Theorem~\ref{TheoPaul} is always smaller or equivalent to System~\ref{Sysys}.  \\

% Thus, $L_\epsilon(D,a)$ of the Theorem~\ref{TheoPaul} allow us to describe a smaller system to compute
% $\mathcal{C}$ the $\epsilon$-cover-template-set for $D$.

%    {\color{red}
%Theorem 33 indique : intesection -> systeme -> résoudre un systeme particulier q
Theorem~\ref{TheoPaul} indicates that determining the $\epsilon$-cover-template-set for $D$ (which corresponds to an intersection of $|D|$ balls in $\mathbb{F}^2_n$ can be reduced to solving a potentially small linear system.
While the resolution of the aforementioned system can be done with powerful
tools (like GUROBI~\cite{pedroso2011optimization}), we deem that simpler
algorithms should be used in this case.
In particular, according to the configuration of $D$, it is possible to obtain a such system linear that it is straightforward to determine the space of the potential solutions and to find a solution with any Markovian scanning algorithm.
More precisely, if $\mathcal{N}$ denotes the set of the possible solutions $N$ for the linear system described in Theorem~\ref{TheoPaul}, we have:
\begin{equation*}
    \label{eq:N_set}
    \mathcal{N} = \prod_{k=1}^{|I|} \{ 0, \dots, \min (\epsilon,|K_k|)\}
\end{equation*}
since, for $k \in \{1, \dots, |I|\}$, $n_{v,k}$ corresponds to the
distance $d_{K_k}(v_k,v)$, which can not be greater than $|K_k|$, and in
the other hand if $d_{K_k}(v_k,v) > \epsilon$ then, $N$ does not belong
to $L$.  One can then be aware that depending on the
dimension of $\mathcal{N}$, finding a solution $N$ can be efficiently
done via either a brute force algorithm in case of small dimensional
set $\mathcal{N}$, or via a more parsimonious algorithm if the
dimension is high.  As the dimension of $\mathcal{N}$ depends among
other factors on $D$, we consider that the use of one of the both
approaches should be determined with regards to pratical
context-specific consideration.  In this paper, we only describe an
algorithm to use in case of high dimensional $\mathcal{N}$ set. We
propose to rely on an efficient and simple algorithm: the Simulating
Annealing algorithm \cite{SANN}.  Nevertheless, even if we illustrate the proposed
methodology with this algorithm, it could be replaced by any
optimization algorithm based on scanning the space.
%In Appendix~\ref{SNNAlgo}, we detail features of Simulating Annealing
Below we detail features of Simulating Annealing algorithm
that we tune in order to obtain good performances in our
numerical study. It is composed of the following parameters:
\begin{itemize}
    \item \textit{Energy:} We define the following energy so that larger it is, the closer $N$ is to solve the linear system:
          \begin{equation}
              \label{eq:SANN_energy}
              E(N) = \sum\limits_{i=1}^{|I|} f((\epsilon - d(v) - AN)_i)
          \end{equation}
          where $f$ is a ReLU type function: $f(x) = \min(0,x)$.
    \item \textit{Cooling Schedule:} In practice, we observe that finding a solution is not sensitive
          to the cooling of the system, see Section~\ref{Temp_recuit}. Then, we propose to choose a linear decreasing
          temperature. The starting temperature is fixed so that at the initial iteration, all potential move must be
          accepted, whatever the chosen initial point is.
    \item \textit{Proposal distribution:} According to computational considerations and for the sake of numerical
          performance, we define a proposal distribution for which the support is the neighbors set. Moreover, we choose
          a non-symmetric proposal that preferentially promotes neighbors that increases the energy~\eqref{eq:SANN_energy}.
    \item \textit{Termination:} The algorithm is terminated either it reaches the maximum iteration number
          (about 200$k$ iterations), or if a solution is found, which corresponds to a vector $N$ with a null energy.
\end{itemize}
%    }
The experimentations of this part are presented  in Section~\ref{TestalgoPaul}.

\subsection{Database Partitioning Algorithm}
\label{DPA}
Using the developments of the sections~\ref{AC} and~\ref{FMT}, we propose
Algorithm~\ref{SpacePartitioningAlgorithm} to partition the template
database. It takes as inputs $D$ a template database and  a
threshold $\epsilon$ and returns an $\epsilon$-MTS.

\begin{algorithm}[]
    \caption{Database partitioning algorithm}
    \label{SpacePartitioningAlgorithm}
    \KwData{$D,\epsilon$}
    \KwResult{$\mathrm{MTS}$}
    Set $s$ to $2\epsilon$. \\
    Set $\mathrm{MTS}$ to $[\ ]$.\\
    \While{$D \not = \emptyset$}{
        Compute cluster $Cls$ using $D$ and $s$.\\
        \ForEach{cluster $c$ in $Cls$}{
            Search the cover template $t$ for $c$.\\
            \If{a cover template $t$ is found for $c \in C$}{
                Set $D$ to  $D\backslash c$ and add $t$ to $\mathrm{MTS}$.\\}
            Set $s$ to $s-1$.\\}}
    \Return{$MTS$.}
\end{algorithm}

%% file: Create_mp.tex
\section{Attack Scenario, Countermeasure and Case Studies}
\label{FMPCBS}

The aim of this section is to show that the method described
Section~\ref{SPM} eases
the computation of a masterkey-set or a master-feature-set. Their computations
are straightforward in the absence of BTP scheme and are still possible if an
invertible transformation is employed, like Biohashing or some other salting
transformations.
Moreover, that kind of attack is analyzed, and a security bound is established
in Section~\ref{SecureParamK}.
% The aim of this part is given a feature vector $v$ or a token $p$ and a template $t$, find
% a token $p$ or a feature vector $v$ such that $\mathcal{T}(v,p)=t$. Then, with
% function in~\ref{GC}, the utility of Part~\ref{SPM} makes sense.
% Thus, in this part,  biometric authentication systems are analysed
% and more particularly the transformation step.

\subsection{Attack Scenario}
\label{GC}
Consider a pair of functions $\mathcal{T}_1^{-1}$ and $\mathcal{T}_2^{-1}$ defined as follows :
\begin{defi}[Token transformation inversion function]
    \label{TTIF}
    The \emph{token tr\-ansformation inversion function} denoted by $\mathcal{T}_1^{-1}$ takes
    $v \in \mathcal{M_F}$ a feature vector and $t \in \Omega$ a template and
    gives $p$ a token such that $\mathcal{T}(v,p)=t$.
\end{defi}

\begin{defi}[Feature transformation inversion function]
    \label{FTIF}
    The \emph{token tr\-ansformation inversion function} denoted by $\mathcal{T}_2^{-1}$ takes
    $p$ a token and $t \in \Omega$ a template and
    gives $v \in \mathcal{M}_F$ a feature vector such that $\mathcal{T}(v,p)=t$.
\end{defi}
Note that we focus on frameworks for which $\mathcal{T}_1^{-1}$ and
$\mathcal{T}_2^{-1}$ can be computed in a reasonable time: at least
linear and at most subexponential. These functions must be
determined case-by-case according to the used biometric transformation. Furthermore,
an attacker seeking to create a master-feature-set (resp. a masterkey-set) can
do it
using $k$ calls to the inverse transformation function
$\mathcal{T}_1^{-1}$ (resp. $\mathcal{T}_2^{-1}$), where $k$ is the number of
templates.
However, the method developed in Section~\ref{SPM} can be used to reduce
the computation complexity.
Actually, the attacker can compute a master-feature-set or a masterkey-set in
only
$\ell$ step with $\ell \leq k$, where $l$ is the number of
clusters.

\subsection{Countermeasure: Managing the Database Size}
\label{SecureParamK}
Consider a biometric system set with a template space of size $n$ and
a threshold~$\epsilon$. Moreover, suppose that the biometric system is
unbiased i.e., each template is randomly chosen in the template space.
There exists a maximum size for a database at $n$ and $\epsilon$ fixed
which minimizes the gain of an attacker with the method presented in
Section~\ref{SPM} and which maximizes the size of that database.
Notice that the following approach can be applied to any biometric
system.

\subsubsection{Prevent an Advantage}
An advantage of an attacker is significant when our database
partitioning method (Section~\ref{SPM}) reduces the complexity of the
initial attack by at least one. Let $k$ be the number of clients
allowed in a database and, $\mathbb{F}_2^n$ the template space. If $k
    \ge \left\lceil2^n \big /\sum\limits_{i=0}^\epsilon
    \binom{n}{i}\right\rceil$ then, there is at least one cluster
containing two or more templates, according to the Dirichlet's box
principle. % \ref{pigeon}.
% \begin{lem}[Pigeonhole principle]
%     \label{pigeon}
%     If $k$ templates are randomly put into $c$ clusters with
%     uniform probability $1/c$, then at least one cluster
%     will hold more than one template if $k > c$.
% \end{lem}
In our case, $c$ is at most: $\left\lceil2^n \big /\sum\limits_{i=0}^\epsilon \binom{n}{i}\right\rceil$
and there are two scenarios:
\begin{enumerate}
    \item There are enough clients to find a coverage of $\mathbb{F}_2^n$ by using their clusters and
          any other enrollment is already compromised.
    \item There is not enough clients to find a coverage of $\mathbb{F}_2^n$ and the attacker obtains an
          advantage for the computation.
\end{enumerate}

By using birthday problem, more particularly the probability of a near
collision~\cite{Lamberger2012,lamberger2012memoryless}, we can establish that,
the average number of template must be about $2^{(n+1)/2}S_\epsilon(n)^{-1/2}$ so that a cluster contained two templates,
where, $\sum\limits_{i=0}^{\epsilon} \binom{n}{i} = S_\epsilon(n)$.
Furthermore, the number of near collisions is $N_C(\epsilon)$ and its expected value $\mathbb{E}(N_C(\epsilon))$ is
equal to $\binom{k}{2}S_\epsilon(n)2^{-n}$ with $k$ the number of templates.
Thus, the number $k$ of templates which give a collision with a probability of $50\%$
is
\begin{equation}
    \label{formula}
    \approx 2^{n/2}S_\epsilon(n)^{-1/2}
\end{equation}

%Table~\ref{ChooseK} and

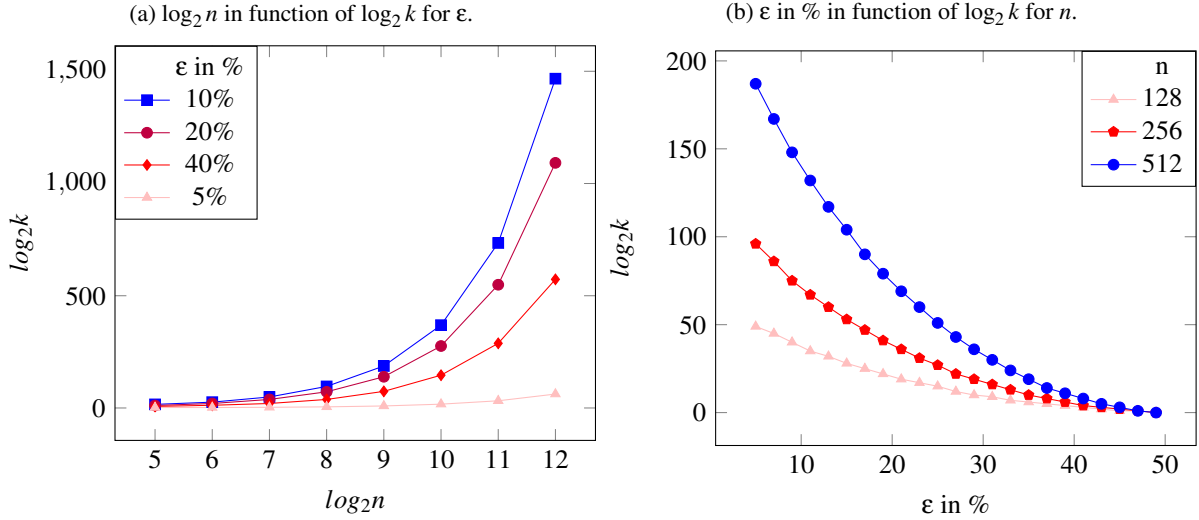
\begin{figure*}[tb]
    \begin{center}
        \begin{subfigure}[b]{0.5\textwidth}
            \subcaption{$\log_2 n$ in function of $\log_2 k$ for  $\epsilon$.}
            \begin{tikzpicture}
                \begin{axis}[width=\textwidth,xtick=data, x tick label style={font=\normalsize},
                        legend style={at={(0,1)},anchor=north west},
                        ylabel={$log_2 k$},xlabel={$log_2 n$}]
                    \addlegendentry{$\epsilon$ in \%}
                    \addlegendimage{empty legend}

                    \addplot [mark=square*, color = blue] table [x=a, y=b, col sep=comma, mark=+] {Mydata.csv};
                    \addlegendentry{$10\%$}
                    \addplot [mark=*,color=purple] table [x=a, y=c, col sep=comma] {Mydata.csv};
                    \addlegendentry{$20\%$}
                    \addplot [mark=diamond*, color= red] table [x=a, y=d, col sep=comma] {Mydata.csv};
                    \addlegendentry{$40\%$}
                    \addplot [mark=triangle*, color = pink]  table [x=a, y=e, col sep=comma] {Mydata.csv};
                    \addlegendentry{$5\%$}
                \end{axis}
            \end{tikzpicture}
        \end{subfigure}%
        \begin{subfigure}[b]{0.5\textwidth}
            \subcaption{$\epsilon$ in \% in function of $\log_2 k$ for  $n$.}
            \begin{tikzpicture}
                \begin{axis}[width=\textwidth, xtick={0,10,20,30,40,50}, x tick label style={font=\normalsize},
                        legend style={at={(1,1)},anchor=north east},
                        ylabel={$log_2 k$},xlabel={$\epsilon$ in \%}]

                    \addlegendentry{n}
                    \addlegendimage{empty legend}

                    \addplot [mark=triangle*, color=pink] table [x=a, y=b, col sep=comma] {Mydata2.csv};
                    \addlegendentry{$128$}
                    \addplot [mark=pentagon*,color=red] table [x=a, y=c, col sep=comma] {Mydata2.csv};
                    \addlegendentry{$256$}
                    \addplot [mark=*,color=blue] table [x=a, y=d, col sep=comma] {Mydata2.csv};
                    \addlegendentry{$512$}
                \end{axis}
            \end{tikzpicture}
        \end{subfigure}
        \caption{Link between $k$ and $n$ or $\epsilon$.}
        \label{graph}
    \end{center}
\end{figure*}

Figure~\ref{graph} provides numerical and graphical
representations based on
experimentations, enlightening on how $k$ behaves relatively to $n$ and $\epsilon$. They show that the size of
a database which can provide collisions is wide smaller than the size of $n$. Furthermore, if $\epsilon$ is
bigger than $20\%$ of $n$, this size dramatically decreases. To keep enough room in a safe database, $n$ must be larger
than $512$ and $\epsilon$ smaller than $51$.

\subsection{Case Study: Addition of Users in the Database}
\label{CS:ADD}

Let $D$ be a database and $C$ a master-template-set for $D$ with
respect to  a threshold $\epsilon$.  As the proposed method of
Section~\ref{SPM} works incrementally, the clustering can be repeated
each time there is a new arrival in the database. However, considering
the cost of such an operation, we propose a more efficient algorithm
yielding a near-optimal solution.  Let us denote by $t$ the new
enrolled template. The steps for adding $t$ are as follows:
\begin{enumerate}
    \item Compute the distance $d_i$ between each center $c_i \in C$ and $t$.
    \item If there exists $d_i \leq \epsilon$ then nothing to do,
          \textit{i.e.,} $t$ has a representative.
    \item Otherwise, add the singleton master-template-set $t$ to $C$.
\end{enumerate}

To minimize $C$ once a certain number of clients have been added,
the partitioning algorithm of Section~\ref{SPM} is applied on $C$ and $D$, and
the smallest resulting set is kept.
Note however that since its computation is faster on $C$ than on $D$
due to its smaller cardinal, it is better to start with $C$.
%and use it until the computation of $D$ is achieved.

\subsection{Case Study: Removing a User from the Database}
\label{CS:REM}
Removing a user from the database is a more complex problem.
Indeed, if the template $t_u$ of a user $u$ is removed from the database, that
may mean that
she has unsubscribed or she has been banned from the service. In
any case, all
the templates in her ball
must be systematically rejected. Thus, there is a problem if the templates $t_{u'}$ of
other users $u'$ are such that $B_{u'} \cap B_0 \not = \emptyset$ with
$B_0=B(t_u,\epsilon)$ and
$B_{t_{u'}}=B(t_{u'},\epsilon)$.
If the number of templates in the database plus the number of removed templates is at most
$  2^{n/2}S_\epsilon(n)^{-1/2}$, this case happens with a small probability.
In this case, all users
have an increased False Rejection Rate (FRR)  proportionally to the
intersection between their ball and $B_0$.
Then, as this case is uncommon,
for the comfort of the users and to keep the initial FRR, it is recommended to re-enroll the affected user(s).
Note that if a large amount of persons are unsubscribed from the database, the
size of the template space is reduced, precisely
by $\sum_{i=0}^\epsilon \binom{n}{i}$ for each removed user.
% Then, to prevent that kind of event, we recommend
% taking a greater $n$ in order of $2\times n$. 
Furthermore, when the number of removed clients $k_1$ and the
number of clients in the database $k_2$ are such that $k_1+k_2 \approx 2^{n/2}S_\epsilon(n)^{-1/2}$, the
system must be changed, or no more users should be accepted as recommended in
Section~\ref{SecureParamK}.

%% file: Attack_eval.tex
\section{Attack Evaluation}
\label{AttackEval}
In this section we provide some experimental evaluations of
Algorithm~\ref{SpacePartitioningAlgorithm} and, we discuss
our results. In our experiments, the passwords are assumed 
unique for each individual. The hashed passwords serve as seeds for the generation of the matrices. Thus, the produced templates are uniformly distributed.

\begin{table*}[tb]
    \begin{center}
        %\resizebox{\textwidth}{!}{%
        \begin{tabular}{|c|c|c|c|c|c|c|c|}
            \hline
            $n$    & $\epsilon $      & $\#$clients       & $\#$clust             & $\#$clust(G) & Efficiency                        & Time (ms)                  & Time G (ms)
            \\ \hline

            $ 15 $ & $  $             & $  $              & $\phantom{0}1.000$    & $27.352$     & $\mathbf{\times 27.35}$           & $1\,239.629$               & $ \phantom{0}7.182 $  \\
            $ 20 $ & $  $             & $  $              & $\phantom{0}2.700$    & $35.433$     & $\mathbf{\times 13.12}$           & $8\,415.270$               & $ 10.714 $            \\
            $ 25 $ & $  $             & $  $              & $\phantom{0}5.260$    & $44.965$     & $\times \phantom{0}8.55$          & $7\,085.071$               & $ 16.302 $            \\
            $ 30 $ & $10$             & $50$              & $\phantom{0}8.709$    & $48.977$     & $\times \phantom{0}5.70$          & $8\,775.802$               & $ 18.940 $            \\
            $ 35 $ & $  $             & $  $              & $12.838$              & $49.872$     & $\times \phantom{0}3.90$          & $6\,870.666$               & $ 21.636 $            \\
            $ 40 $ & $  $             & $  $              & $18.087$              & $49.986$     & $\times \phantom{0}2.77$          & $ 6\,417.596 $             & $ 23.762 $            \\
            $ 45 $ & $  $             & $  $              & $23.217$              & $49.998$     & $\times \phantom{0}2.14$          & $ 6\,773.971 $             & $ 25.629 $            \\
            \hline
            \hline
            $  $   & $ \phantom{0}3 $ & $  $              & $200.000$             & $ 200.000$   & $ \times \phantom{0}1.00$         & $ \phantom{0000}42.252 $   & $ 558.165 $           \\
            $  $   & $ \phantom{0}5 $ & $  $              & $200.000$             & $ 200.000$   & $ \times \phantom{0}1.00$         & $ \phantom{0000}43.969 $   & $ 449.166 $           \\
            $  $   & $ 10 $           & $  $              & $198.280$             & $ 200.000$   & $ \times \phantom{0}1.00$         & $ \phantom{000}799.243 $   & $ 329.992 $           \\
            $70$   & $ 15 $           & $200$             & $\phantom{0}90.000$   & $ 200.000$   & $ \times \phantom{0}2.22$         & $ \phantom{0}47\,016.050 $ & $ 337.082 $           \\
            $  $   & $ 20 $           & $  $              & $\phantom{0}46.543$   & $199.985$    & $ \times \phantom{0}4.30$         & $ \phantom{0}94\,364.038 $ & $ 350.846 $           \\
            $  $   & $ 25 $           & $  $              & $\phantom{0}22.109$   & $198.982$    & $\mathbf{\times \phantom{0}9.00}$ & $ 222\,386.614 $           & $ 346.420 $           \\
            $  $   & $ 30 $           & $  $              & $ \phantom{0}10.000 $ & $182.891$    & $\mathbf{\times 18.29}$           & $ \phantom{0}37\,925.217 $ & $ 326.385 $           \\
            $  $   & $ 35 $           & $  $              & $ \phantom{00}3.600 $ & $137.583$    & $\mathbf{\times 38.20}$           & $ 949\,691.855 $           & $ 380.933 $           \\
            \hline
            \hline
            $  $   & $  $             & $ \phantom{0}30 $ & $\phantom{0}29.96$    & $30$         &                                   & $24.318 $                  & $ \phantom{0}16.446 $ \\
            $  $   & $  $             & $ \phantom{0}50 $ & $\phantom{0}49.83$    & $50$         &                                   & $46.362 $                  & $ \phantom{0}40.018 $ \\
            $  $   & $  $             & $ \phantom{0}70 $ & $\phantom{0}69.78$    & $70$         &                                   & $78.995 $                  & $ \phantom{0}80.496 $ \\
            $  $   & $  $             & $ \phantom{0}90 $ & $\phantom{0}89.67$    & $90$         &                                   & $136.572 $                 & $ 137.186 $           \\
            $50$   & $10$             & $ 110 $           & $109.60$              & $110$        & $ \times 1$                       & $208.685 $                 & $ 189.203  $          \\
            $  $   & $  $             & $ 130 $           & $129.30$              & $130$        &                                   & $428.885 $                 & $ 251.221 $           \\
            $  $   & $  $             & $ 150 $           & $148.93$              & $150$        &                                   & $670.493 $                 & $ 335.790 $           \\
            $  $   & $  $             & $ 170 $           & $168.79$              & $170$        &                                   & $531.363 $                 & $ 434.727  $          \\
            $  $   & $  $             & $ 190 $           & $188.46$              & $190$        &                                   & $625.553 $                 & $ 562.672 $           \\
            \hline
        \end{tabular}
        %}
    \end{center}
    \caption{Summary of the experiments of the space partitioning algorithm.}
    \label{Exp1}
\end{table*}

% \begin{rem}
%     When $\epsilon \ll n$ the gain are less significant.
% \end{rem}
% 
% \begin{rem}
%     When "Nbr clients" grow up the gain are more interesting.
% \end{rem}

\begin{table*}[tb]
    \begin{center}
        %\resizebox{\textwidth}{!}{%
        \begin{tabular}{|c|c|c|c|}
            \hline
            $n$    & $\epsilon $ & $\#$clients & Time (ms)    \\
            \hline
            $ 15 $ &             &             & $1\,277.473$ \\
            $ 20 $ &             &             & $1\,592.213$ \\
            $ 25 $ &             &             & $2\,033.807$ \\
            $ 30 $ & $10$        & $ 50 $      & $2\,428.682$ \\
            $ 35 $ &             &             & $2\,758.428$ \\
            $ 40 $ &             &             & $3\,887.738$ \\
            $ 45 $ &             &             & $3\,866.077$ \\
            \hline
        \end{tabular}
        \begin{tabular}{|c|c|c|c|}
            \hline
            $n$   & $\epsilon $    & $\#$clients & Time (ms)      \\
            \hline
                  & $\phantom{0}3$ &             & $24\,901.480 $ \\
                  & $\phantom{0}5$ &             & $24\,949.724 $ \\
                  & $10 $          &             & $22\,279.933 $ \\
            $70 $ & $15 $          & $200 $      & $20\,978.806 $ \\
                  & $20 $          &             & $21\,028.472 $ \\
                  & $25 $          &             & $29\,089.280 $ \\
                  & $30 $          &             & $31\,586.784 $ \\
            \hline
        \end{tabular}
        \begin{tabular}{|c|c|c|c|}
            \hline
            $n$    & $\epsilon $ & $\#$clients       & Time (ms)                 \\
            \hline
                   &             & $ \phantom{0}70 $ & $ \phantom{0}8\,474.519 $ \\
                   &             & $ \phantom{0}90 $ & $ 11\,087.893 $           \\
                   &             & $ 110 $           & $ 17\,464.348 $           \\
            $ 70 $ & $ 10 $      & $ 130 $           & $ 18\,330.508 $           \\
                   &             & $ 150 $           & $ 19\,009.142 $           \\
                   &             & $ 170 $           & $ 20\,887.950 $           \\
                   &             & $ 190 $           & $ 19\,539.516 $           \\
            \hline
        \end{tabular}
        %}
    \end{center}
    \caption{Summary of the experiments of the $\epsilon$-cover-template search algorithm ILP version.}
    \label{Exp2}
\end{table*}

\subsection{Naive Greedy Approach}
\label{Gluttony}
To compare the efficiency of our proposal with a baseline, we propose a naive
algorithm based on
a greedy strategy. First, a template is picked from the template database.
Then, all templates in the template database which are at a distance of at most
$\epsilon$ from the chosen template are removed.
%{\color{red} (XXX distance par rapport a qui ? XXX)}. 
These steps are repeated as long as there are templates in the database.
As a result, the chosen templates form the MTS.
% {\color{red} The proposed algorithm is detailed in the appendices. (Est-ce 
% vraiment utile ?)}

\subsection{Evaluation of the Database Partitioning}
\label{SPEval}

% {\color{red} Rmqs :

% - Améliorer le titre de la section

% - Faire attention aux lettres capitales qui sont mal employées dans les
% référencements

% - Trouver un moyen de faciliter la lecture des notations, un tableau p-e ?

% - Le ``Gain VS G'' un peu étrange (normalement, un algo vs un algo). C'est donc
% VS qui est mal employé, d'où peut-être le choix d'un nom (XX vs Greedy ?)

% - Dans la table 1, à la place de Gain vs G, peut-être qu'on peut employer le
% terme speedup (par rapport à la baseline).

% - Dans les tables 3 et 4, des colonnes contiennent une seule valeur répétée à
% chaque ligne. Il faudrait enlever toutes les répétitions (comme dans les tables
% 1 et 2)

% - Dans tout le papier, les mots/termes suivant apparaissent trop souvent :
% method, moreover, thus.
%}

Templates are randomly drawn from $\mathbb{F}_2^n$.
For each configuration, experimentations are replicated $10\,000$ times and
averaged results are computed.
The average results are presented in Table~\ref{Exp1} and Table~\ref{Exp2} with
the following notations:
\begin{itemize}
    \item $n$: the space dimension.
    \item $\epsilon$: the threshold.
    \item $\#clients$: the number of templates in the TDB.
    \item $\# clust$: the number of clusters found with Algorithm~\ref{SpacePartitioningAlgorithm}.
    \item $\# clust (G)$: the number of clusters found using the greedy Algorithm~\ref{Gluttony}.
    \item $Efficiency$ is the ratio $\# clust (G) / \# clust$.
    \item $Time$ is the running time of Algorithm~\ref{SpacePartitioningAlgorithm}.
    \item $Time (G)$ is the running time of the greedy Algorithm~\ref{Gluttony}.
\end{itemize}
As the computation of the $\epsilon$-cover-template~\ref{FMT} is the
most expensive part of Algorithm~\ref{SpacePartitioningAlgorithm}, an
experimentation Table~\ref{Exp2} is dedicated to the
$\epsilon$-cover-template search~\ref{FMT}. In fact,
we remark that the gain of the attacker is greater when the
value of $k$ is greater that what we recommend in Section~\ref{SecureParamK}.

\subsection{Evaluation of Simulated Annealing}
\label{TestalgoPaul}
Keeping the same notations, the average experimentations are stored in
Table~\ref{Exp4}.  In the case where the simulated annealing is used
as a sub-routine of the algorithm~\ref{SpacePartitioningAlgorithm},
this latter is slower and less efficient. The main reason of this loss
of performance is the error rate of the simulated annealing which
forces doing more calculations. However, it is quicker and more
efficient than solving a system to answer to the near string problem
given in Section~\ref{MCSP}.

% \begin{figure}[ht]
%     \begin{center}
%         \resizebox{\textwidth}{!}{%
%             \begin{tabular}{|c|c|c|c|c|}
%                 \hline
%                 $n$    & $\epsilon $ & $\#$clients & $\#$ clust & Time            \\
%                 \hline
%                 $ 15 $ & $ 10 $      & $ 50 $      & $ 1.991 $  & $ 23760 $ ms    \\
%                 $ 20 $ & $ 10 $      & $ 50 $      & $ 5.54 $   & $ 42458150 $ ms \\
%                 \hline
%             \end{tabular}}
%     \end{center}
%     \caption{Summary of the experiments over space partitioning.}
%     \label{Exp3}
% \end{figure}

\begin{table*}[tb]
    \begin{center}
        \resizebox{\textwidth}{!}{%
            \begin{tabular}{|c|c|c|r|r|}
                \hline
                $n$  & $\epsilon $ & $\#$clients & Error   & Time   \\
                     &             &             & in \%   & (ms)   \\
                \hline
                $15$ &             &             & $16.34$ & $1201$ \\
                $20$ &             &             & $0.64$  & $17$   \\
                $25$ &             &             & $0.00$  & $1$    \\
                $30$ & $10$        & $50$        & $0.00$  & $1$    \\
                $35$ &             &             & $0.00$  & $1$    \\
                $40$ &             &             & $0.05$  & $1$    \\
                $45$ &             &             & $0.36$  & $2$    \\
                \hline
            \end{tabular}
            \begin{tabular}{|c|r|c|r|c|}
                \hline
                $n$  & $\epsilon $ & $\#$clients & Error  & Time \\
                     &             &             & in \%  & (ms) \\
                \hline
                     & $3$         &             & $0.15$ & $36$ \\
                     & $5$         &             & $0.00$ & $36$ \\
                     & $10$        &             & $0.00$ & $35$ \\
                $70$ & $15$        & $200$       & $0.00$ & $36$ \\
                     & $20$        &             & $0.00$ & $37$ \\
                     & $25$        &             & $0.00$ & $40$ \\
                     & $30$        &             & $0.00$ & $40$ \\
                \hline
            \end{tabular}
            \begin{tabular}{|c|c|c|c|r|}
                \hline
                $n$  & $\epsilon $ & $\#$clients     & Error  & Time \\
                     &             &                 & in \%  & (ms) \\
                \hline
                     &             & $\phantom{0}70$ & $1.95$ & $6$  \\
                     &             & $\phantom{0}90$ & $0.14$ & $12$ \\
                     &             & $110$           & $0.00$ & $18$ \\
                $70$ & $10$        & $130$           & $0.00$ & $22$ \\
                     &             & $150$           & $0.00$ & $27$ \\
                     &             & $170$           & $0.00$ & $31$ \\
                     &             & $190$           & $0.00$ & $34$ \\
                \hline
            \end{tabular}
        }
    \end{center}
    \caption{Summary of the experiments of the $\epsilon$-cover-template search algorithm SANN version.}
    \label{Exp4}
\end{table*}

\label{Temp_recuit}

Moreover, we use several cooling functions (from~\cite{SNNLog,SANN})
to determine what is preferable.  We draw randomly $10\,000$ times
some templates in a ball of radius $\epsilon$ and a template as a
center. These templates except the center form the simulated database.
By doing so, we are sure that the database enables an
$\epsilon$-cover-template. Then, for each database generated, an
$\epsilon$-cover-template is sought. The results of Table~\ref{Exp5}
in Appendix indicate that finding a solution is not
strongly sensitive to the cooling method of the system.  However, all
cooling methods give similar results.

% \subsection{Biohashing Attack Evaluation}
% 
% 
% For each configuration, experimentation are replicate $1000$ times and
% an average is computed. Templates are drawn randomly in $\mathbb{F}_2^n$ and the feature
% vector is drawn randomly in $[0;1]^m$ with $m < n$.
% The average experimentation are stored in ~\ref{BAE}
% 
% \begin{figure}[ht]
%     \begin{center}
%         \resizebox{.47\textwidth}{!}{%
%             \begin{tabular}{|c|c|c|}
%                 \hline
%                 Size of template & Size of feature vector & Time      \\
%                 $4$              & $5$                    & $0.7$ s   \\
%                 $5$              & $7$                    & $1.9$ s   \\
%                 $7$              & $10$                   & $151$ s   \\
%                 $9$              & $12$                   & $>3500$ s \\
%                 \hline
%             \end{tabular}}
%     \end{center}
%     \caption{Biohashing attack evaluation table}
%     \label{BAE}
% \end{figure}
% 
% As the experimentation shows that the algorithm is exponential, the experimentation
% have not been pursued.

%% file: Conclusion.tex
\section{Concluding Remarks}\label{Concl}

%{\color{red} Suggestion : faire des paragraphes.}
%{\color{red} Maintenant les paragraphes sont trop courts !}

In this paper, we have performed an in-depth analysis of the Hamming
space as template space.  We first have introduced some formal
definitions such as multiplicative near-collision, master-template,
$\epsilon$-covering template and some technical terms and concepts. We
then have proposed an algorithm to perform a partition of the set of
templates.  This partitioning can be used to improve either the
masterkey-set search or the master-feature-set search. The proposed
center search algorithm using simulated annealing is also a result of
independent interest for solving the near-string-problem.  If an
additional and secure authentication factor is put into place, this
partitioning may be used in the authentication mode to compress the
template database. This compression is achieved by replacing each
template by a reference of its cluster center.

By relying on the properties of near-collisions and the partitioning algorithm, 
we
have also shown there exists a security bound on the size of a
database that depends on both the space dimension and the decision threshold.
Beyond that limit on the size, there is a high probability
of a near collision that impacts both security and efficiency.
Identification and authentication systems are negatively affected
by near-collisions, especially identification in terms of recognition accuracy.
%Thus, we used 
%near-collisions
%and the previous algorithm to set a security parameter on the size of the 
%databases.
% We set a limit for database such that after that limit, there is a high probability
% of a near collision and decrease both security and efficiency.
A countermeasure when the number of templates exceeds the recommended security
parameter is to extend the size of the template space.
This can be transparently done for the user if the risk has been anticipated in
the design of the protocol. For instance, in the case of projection-based
transformations, the enrollment template can easily be extended using a part of
an extended fresh template.
% Additionally, we have proposed an algorithm of independent 
% interest to solve the
% near-string-problem using simulated annealing.

% Some formal definitions such as multiplicative
% near-collision, master-template, $\epsilon$-covering template and, some 
% technical terms and
% concepts have been introduced.

%% file: Annexes.tex
%\section{Appendices}
\appendix

\section{Omitted Proofs}

%\subsubsection{Proof of Theorem~\ref{MYNPHARD}}

% {\color{red} Rmq : Remplacer l'opérateur/symbole $\prec_{\mathcal{P}}$ par une 
% phrase}

\begin{proof}[Proof of Theorem~\ref{MYNPHARD}]\label{MYNPHARDProof}
    Let $A$ be the oracle for the modified closest-string problem and ($S$) a closest-string problem instance.
    Thus, on at most $n$ calls to $A$, the closest-string problem can be solved. In fact, the solver $B$
    of the closest-string problem sends to $A$ the following instances: $(S,1),(S,2),\dots,(S,i\leq n)$ and stops
    at the $i$-th instance for which $A$ finally comes with the solution $t$.
    Then, $B$ returns the pair ($t,i$), solution to the initial problem.
    Since $B$ can be reduced in polynomial time to $A$ and $B$ is $NP$-hard,
    $A$ is also $NP$-hard.
    The reduction is trivial in the other direction. $\qed$
\end{proof}

%\subsubsection{Proof of Theorem~\ref{Intertheo}}

\begin{proof}[Proof of Theorem~\ref{Intertheo}]\label{Intertheoproof}
    Let $p \in \cap_{u \in D} B(u,\epsilon)$. Then, $\forall u \in D, p\in B(u,\epsilon)$. Which implied that
    $\forall u \in D, d_H(p,u) \leq \epsilon$.
    And so, $p$ is an $\epsilon$-cover-template for $D$. Then, $p \in \mathcal{C} $
    which implies that $ \cap_{u \in D} B(u,\epsilon) \subset \mathcal{C}$.
    Moreover, let $p \in C$. Then, $\forall u \in D, d_H(p,u) \leq \epsilon$. So, $\forall u \in D, p\in B(u,\epsilon)$.
    Thus, $P \in \cap_{u \in D} B(u,\epsilon)$ and, $\mathcal{C} \subset \cap_{u \in D} B(u,\epsilon)$.
    Then, using both inclusion, $\mathcal{C} = \cap_{u \in D} B(u,\epsilon)$. $\qed$
\end{proof}

%\subsubsection{Lemma~\ref{LemmaPaul} and Proof}

\begin{proof}[Proof of Lemma~\ref{LemmaPaul}]\label{LemmaPaulProof}
    Let $D\subset \mathbb{F}_2^n$ be a template database such that $| D | > 1$
    and $K_i=\lbrace i \rbrace, \forall i \in \lbrace 1,\dots,n\rbrace$.
    Therefore, $\sqcup_{i \in \left\lbrace 1, \dots,n \right\rbrace} K_i = \left\lbrace 1, \dots,n \right\rbrace$ and,
    $\forall i \in \lbrace 1,\dots,n\rbrace$, $\mathcal{P}_D(K_i)= \text{True}$. Then,
    $I = \lbrace K_i, \forall i \in \left\lbrace 1, \dots,n \right\rbrace\rbrace
        \not = \emptyset$. $\qed$
\end{proof}

%\subsubsection{Proof of Theorem~\ref{TheoPaul}}
% {\color{red} Rmq : Revoir la rédaction des preuves suivantes :}

\begin{proof}[Proof of Theorem~\ref{TheoPaul}]\label{TheoPaulProof}
    Let $D$ be a database, $u \in D$ a template, $p\in \mathbb{F}_2^n$ a template
    and, $A_K(u)= \lbrace v \in D | d_K(u,v) = 0 \rbrace$.
    There are two cases:
    \begin{enumerate}
        \item If $v \in A_K(u)$ then, $d_k(u,v)=0$.
        \item Else, $v \in A_K(u)^c$ then, $d_k(u,v)= | K |$.
    \end{enumerate}
    If $v \in A_K(u)^c$ then, $d_K(v,p) = | K | - d_K(p,u)$.
    However, as $I$ is a partition of $\lbrace 1, \dots,n \rbrace$,
    $d_H(u,p) = \sum \limits_{K \in I} d_K(u,p)$.

    Suppose that $p \in C$ the
    $\epsilon$-cover-template-set for $D$.
    As $max_{u \in D} d_H(u,p) \leq \epsilon$ then, $\sum \limits_{K \in I} d_K(u,p) \leq \epsilon$.
    Thus, for $v \in D$, $d_K(v,p) = d_K(p,u)\mathbb{1}_{A_K(u)}(v) + (| K | - d_K(u,p))\mathbb{1}_{A_K(u)^c}(v)$.
    Then, for a given couple $(u,v)$, we have:
    \begin{eqnarray*}
        \sum\limits_{K\in I} d(v,p) &=& \sum\limits_{K\in I} d_K(p,u)\mathbb{1}_{A_K(u)}(v)\\
        & &+ (| K | - d_K(u,v))\mathbb{1}_{A_K(u)^c}(v) \\
        &=& \left(\sum\limits_{K\in I} d_K(p,u) (\mathbb{1}_{A_K(u)}(v) - \mathbb{1}_{A_K(u)^c}(v))\right)\\
        & &+  \sum\limits_{K\in I} | K | \mathbb{1}_{A_K(u)^c}(v) \\
    \end{eqnarray*}
    Moreover, $d_H(u,v) = \sum\limits_{K\in I} | K | \mathbb{1}_{A_K(u)^c}(v)$ then,
    $$\sum\limits_{K\in I} d(v,p) = \sum\limits_{K\in I} d_K(p,u) (\mathbb{1}_{A_K(u)}(v) - \mathbb{1}_{A_K(u)^c}(v)) + d_H(u,v).$$
    Then,
    \begin{eqnarray*}
        &\sum \limits_{K \in I} d_K(v,p) \leq \epsilon \\
        \Leftrightarrow & \sum\limits_{K\in I} d_K(p,u) (\mathbb{1}_{A_K(u)}(v) - \mathbb{1}_{A_K(u)^c}(v)) \leq \epsilon- d_H(u,v)\\
        \Leftrightarrow & A(u)d_K(p,u) \leq \epsilon- d_H(u,v) \\
        \Leftrightarrow & p \in L
    \end{eqnarray*}
    $\qed$
\end{proof}

\section{Illustration for the Partitioning of the Set of Indices}
\label{EXFMT}

The following example serves as an illustration for  Section~\ref{FMT}.
Let $D = \lbrace (1, 0, 1, 1, 0, 1, 1),$ $(1, 0, 0, 1, 0, 1, 1), (1, 0, 1, 1, 1,
    1, 1), (1, 0, 0, 1, 1, 1, 0) \rbrace$ be a database represented as a
matrix with the templates in rows. The identical or opposite columns are
labelled
with the same symbol, as follows :

\begin{center}
    \begin{tabular}{|c|c|c|c|c|c|c|c|}
        \hline
              & $1$          & $2$          & $3$            & $4$          & $5$        & $6$          & $7$              \\
        \hline
        $v_1$ & $1$          & $0$          & $1$            & $1$          & $0$        & $1$          & $1$              \\
        $v_2$ & $1$          & $0$          & $0$            & $1$          & $0$        & $1$          & $1$              \\
        $v_3$ & $1$          & $0$          & $1$            & $1$          & $1$        & $1$          & $1$              \\
        $v_4$ & $1$          & $0$          & $0$            & $1$          & $1$        & $1$          & $0$              \\
        \hline
              & $\spadesuit$ & $\spadesuit$ & $\blacksquare$ & $\spadesuit$ & $\bigstar$ & $\spadesuit$ & $\blacktriangle$ \\
        \hline
    \end{tabular}
\end{center}

\noindent
We remind the statement $\mathcal{P}_D(K)$ which holds
if for all templates of $D$, their pairwise distance is $|K|$ or $0$.
Let $K = \lbrace 1,2,4,6 \rbrace$ be the set of columns marked with a
$\spadesuit$. Then, $\mathcal{P}_D(K)$ holds.
However, for $K = \lbrace 3,7 \rbrace$, $\mathcal{P}_D(K)$ does not hold. In fact,
if $K$ is uniquely comprised of columns having the same symbol, the statement holds.
If the columns which are identical or opposite are merged together,
the property $\mathcal{P}_D(K)$ holds. Finally, in
this example, the partition $I$ is
$\lbrace \underbrace{\lbrace 1,2,4,6 \rbrace}_{\spadesuit},\underbrace{\lbrace 3
        \rbrace}_{\blacksquare},\underbrace{\lbrace 5 \rbrace}_{\bigstar}
    ,\underbrace{\lbrace 7 \rbrace}_{\blacktriangle}  \rbrace$.

% {\color{red}Rmq 1 : Le reviewer peut se demander ce qu'il se passe lorsque le
% nombre de templates augmente dans la base de données. Dans cet exemple on
% pourrait croire que les chances que la propriété soit vraie diminue avec le
% nombre de templates.
% Rep : Dans le pire des cas, I est l'union des singletons ce qui arrive avec une forte proba si il n'y a ne serais que un 
% template éloigné des autres. C'est une construction qui a du sens si les templates sont proche. Après si ils sont dans un même
% prérimètre tout va bien et I va être petit.}

\section{Comparison of Cooling Strategies}

Some experimentations with our Simulated Annealing process have been done
using different temperatures
and the same number of steps. The objective is to figure out which
cooling strategy is preferable. Table~\ref{Exp5} indicates
that the different options do not significantly impact
the efficiency of the Simulated Annealing algorithm.

\begin{table}[ht]
    \resizebox{0.48\textwidth}{!}{%
        \begin{tabular}{|c|c|c|c|c|c|}
            \hline
            Cooling method        & $n$  & $\epsilon $ & $\#clients$ & Error   & Time \\
                                  &      &             &             & in $\%$ & (ms) \\
            \hline
                                  & $45$ & $10$        & $50$        & $0.1$   & $7$  \\
                                  & $50$ & $10$        & $50$        & $0.6$   & $11$ \\
            Additive cooling      & $55$ & $10$        & $50$        & $3.1$   & $9$  \\
                                  & $60$ & $10$        & $50$        & $8.3$   & $11$ \\
                                  & $65$ & $10$        & $50$        & $47.2$  & $19$ \\
            \hline
            \hline
                                  & $45$ & $10$        & $50$        & $0.6$   & $12$ \\
                                  & $50$ & $10$        & $50$        & $0.8$   & $11$ \\
            Linear multiplicative & $55$ & $10$        & $50$        & $3.7$   & $5$  \\
                                  & $60$ & $10$        & $50$        & $5.3$   & $16$ \\
                                  & $65$ & $10$        & $50$        & $35.3$  & $18$ \\
            \hline
            \hline
                                  & $45$ & $10$        & $50$        & $0.2$   & $7$  \\
                                  & $50$ & $10$        & $50$        & $1.2$   & $10$ \\
            Exponential           & $55$ & $10$        & $50$        & $4.3$   & $3$  \\
                                  & $60$ & $10$        & $50$        & $6.9$   & $4$  \\
                                  & $65$ & $10$        & $50$        & $40.8$  & $15$ \\
            \hline
            \hline
                                  & $45$ & $10$        & $50$        & $0.5$   & $6$  \\
                                  & $50$ & $10$        & $50$        & $1.4$   & $3$  \\
            Logarithmic           & $55$ & $10$        & $50$        & $2.9$   & $3$  \\
                                  & $60$ & $10$        & $50$        & $6.6$   & $10$ \\
                                  & $65$ & $10$        & $50$        & $40.8$  & $10$ \\
            \hline
        \end{tabular}}
    \caption{Comparison of cooling methods for our simulated annealing.}
    \label{Exp5}
\end{table}